\documentclass[fleqn,twoside]{article}
\usepackage{espcrc2}
\usepackage{natbib}

\usepackage{graphicx}
\usepackage[figuresright]{rotating}

\newcommand{\AmS}{{\protect\the\textfont2
  A\kern-.1667em\lower.5ex\hbox{M}\kern-.125emS}}
\def\lapp{\ifmmode\stackrel{<}{_{\sim}}\else$\stackrel{<}{_{\sim}}$\fi}
\def\gapp{\ifmmode\stackrel{>}{_{\sim}}\else$\stackrel{>}{_{\sim}}$\fi}
\newcommand{\tfe}{1E~1048--5937}
\newcommand{\tfn}{1E~2259+586}

\newcommand{\soe}{RXS~1708--4009}
\newcommand{\oft}{4U~0142+61}
\newcommand{\efo}{1E~1841--045}
\newcommand{\axj}{AX~J1845.0--0258}
\newcommand{\ett}{XTE~J1810--197}
\newcommand{\cxo}{CXOU~J0110043.1--721134}

\title{(Anomalous) X-ray Pulsars}

\author{V. M. Kaspi\address[]{Physics Department, McGill University, Rutherford Physics Building, 3600 University St.\\ Montreal, QC, Canada  H3A 2T8}\thanks{Steacie Fellow, Canada Research Chair},
       {F. P. Gavriil\addressmark}
        }
       
\begin{document}

\begin{abstract}
We review the observational properties of the class of young
neutron stars known as ``anomalous X-ray pulsars,'' emphasizing
the tremendous progress that has been made in recent years, and
explain why these objects, like the ``soft gamma repeaters,'' are today 
thought to be young, isolated, ultrahigh magnetic field neutron
stars, or ``magnetars.''
\vspace{1pc}
\end{abstract}

\maketitle

\section{INTRODUCTION}

Prior to the commissioning of {\it BeppoSAX} and the {\it Rossi X-ray
Timing Observatory} in 1996, the so-called ``Anomalous'' X-ray Pulsars
(AXPs) were considered very mysterious sources, because the energy source
for their bright X-ray emission was unknown.  At the time, there
were only 3 known members of this class.  They were distinguished by
having periods in the narrow range 6--9~s, showing approximately steady
spin-down, and having softer spectra in general that those seen in
accreting X-ray pulsars.  All were known to lie within 1$^{\circ}$ of
the Galactic Plane, and interestingly, one source, 1E~2259+586, was known to
reside in the supernova remnant CTB~109.  AXPs as a class were identified as
having modest X-ray luminosities, in the range $L_x \sim
10^{34}-10^{35}$~erg~s$^{-1}$.  The leading model to explain the AXPs
was that they were accreting neutron stars, though with properties very
different from the bulk of established accreting X-ray pulsars,
including the absence of any evidence of a companion \citep{vtv95,ms95}.

The situation post-{\it BeppoSAX} and especially in the latter years of
{\it RXTE} is very different and much clearer.  The basic phenomenology
of the sources is now well mapped out.  Here, we systematically review the
most important properties of this class of objects, which now includes 5
and possibly 8 sources (see Tables~\ref{ta:axps1} and \ref{ta:axps2}),
and summarize why today, accretion models are strongly disfavored.
Rather, the magnetar model, in which AXPs are isolated
young neutron stars powered by a decaying ultrahigh magnetic field,
provides the most compelling explanation for the unusual AXP source
properties, as it does for an equally as exotic class, the soft gamma
repeaters (SGRs).  AXPs have also been reviewed recently by \citet{mcis02},
and magnetars in general have been reviewed recently by
\citet{kas03a,kas03b}.


%
%
%
%

\begin{sidewaystable}
\caption{Spin parameters for AXPs.}
\label{ta:axps1}
\newcommand{\m}{\hphantom{$-$}}
\newcommand{\cc}[1]{\multicolumn{1}{c}{#1}}
\renewcommand{\arraystretch}{1.2} 
\begin{tabular*}{\textheight}{@{\extracolsep{\fill}}lcccccccc}
\hline
 Source & \cc{Distance$^{\dagger}$}  & \cc{SNR} &  \cc{$P$}  & \cc{$\dot{P}$} & \cc{$B_{dp}$}  & \cc{$\dot{E}_s$}  & \cc{$\tau_c$}  & \cc{Ref.}\\
        &  (kpc)                     &          &   (s)      & ($\times 10^{-11}$) & ($\times 10^{14}$~G) & ($\times 10^{32}$~erg s$^{-1}$) & (kyr) & \\
\hline 
\oft     &  $\gapp 1.0$ or $\gapp2.7$ & $-$     &  8.69 & 0.196 & 1.3 & 1.2 & 7.0   &  1 \\
\tfe     & $\gapp2.7$  & $-$     &  6.45 & $\sim 3.81$     & $\sim 5.0$  & $\sim 55$ & $\sim 2.7$ & 2 \\
\soe     & $\sim8$ & $-$     & 11.00 & 1.86 & 4.6 & 5.4 & 9.4   & 3\\
\efo     &  5.7-8.5 & Kes 73 & 11.77 & 4.16 & 7.1 & 9.9 & 4.5   & 4  \\
\tfn     &  3 & CTB 109 &  6.98 & 0.0483   & 0.59 & 0.55& 230 & 5 \\
\axj$^*$ &  $\sim8$ & Kes 75  &  6.97 & $-$ & $-$ & $-$ & $-$   &  6  \\
\cxo$^*$ & 57 & $-$     &  8.02 & $-$ & $-$ & $-$ & $-$ &  7  \\
\ett$^*$ &  $\sim10$ & $-$     &  5.54 & 1.15 & 2.6 & 26 & 7.6   &  8   \\
\hline
\end{tabular*}\\[2pt]
($*$) not confirmed; ($\dagger$) see   \"Ozel, Psaltis \& Kaspi 2001 for a discussion on distance estimates for the confirmed AXPs; References: (1) Gavriil \& Kaspi 2002; (2) Kaspi et al.~2001; (3) Kaspi \& Gavriil 2003; (4) Gotthelf et al.~2002; (5) Woods et al.~2003; (6) Torii et al.~1998; (7) Lamb et al.~2003; (8) Ibrahim et al.~2003. 
\end{sidewaystable}

\begin{sidewaystable}
\caption{Spectral parameters for AXPs.}
\label{ta:axps2}
\newcommand{\m}{\hphantom{$-$}}
\newcommand{\cc}[1]{\multicolumn{1}{c}{#1}}
\renewcommand{\arraystretch}{1.2} 
\begin{tabular*}{\textheight}{@{\extracolsep{\fill}}lcccccc}
\hline
 Source & \cc{$n_H$}   & \cc{$\Gamma$}    & \cc{$kT$}  & \cc{$L_x$} & \cc{$f_{\mathrm{pl}}$ ($\%$)$^{\dagger}$} & \cc{Ref.} \\
        & ($\times 10^{22}$\ cm$^{-2}$) &   &  (keV) &  (erg s$^{-1}$) & & \\
\hline 
\oft      & 0.88 & 3.3  & 0.42  & $3.3\times 10^{34}$  & $\sim88$  & 1\\
\tfe      & 1.0  & 2.9  & 0.63  & $3.4\times 10^{34}$  & $\sim80$  & 2\\
\soe      & 1.49 & 3.11 & 0.45  & $6.8\times 10^{35}$  & $\sim 73$ & 3\\
\efo      & 2.0  & 2.26 & $-$   & $2.3\times 10^{35}$  & $100$     & 3\\
\tfn      & 0.93 & 3.6  & 0.41  & $1\times 10^{35}$    & $\sim50$  & 4\\
\axj$^*$  & 9.0  & 4.6  & $-$   & $7.4\times 10^{34}$  & $100$     & 5\\
\cxo$^*$  & 0.14 & $-$  & 0.41  & $1.5\times 10^{35}$  & $0$       & 6\\
\ett$^*$  & 1.05 & 3.75 & 0.668 & $1.6\times 10^{36}$  & $\sim70$  & 7\\
\hline
\end{tabular*}\\[2pt]
($*$) not confirmed; ($\dagger$) contribution of the power-law component to the total flux, see Perna et al.~2001 for further discussion; References: (1) Juett et al.~2002; (2) Tiengo et al.~2002; (3) Mereghetti et al.~2002; (4) Patel et al~2001; (5) Torii et al.~1998; (6) Lamb et al.~2003; (7) Ibrahim et al.~2003. 
\end{sidewaystable}


\section{TIMING PROPERTIES OF AXPS}

Since their discovery, AXPs have been known to be spinning down.
Unlike most known accreting X-ray pulsars, no evidence was seen, in
nearly two decades of timing, for any extended spin-up.  However, some
deviations from simple spin-down were observed.  AXP 1E2259+586 showed
a handful of possible very short lived spin-up events
\citep[e.g.][]{bs96} as did 1E~1048$-$5937 \citep[e.g.][]{opmi98}.
These were noted by various authors and were suggested to be due to
accretion torque variations \citep[e.g.][]{bs96}, glitches
\citep{hh99}, and magnetar radiative precession \citep{mel97}.
However, with sparse observations consisting of a frequency measurement
every few years and rarely more often, determining the origin of the
apparent deviations from simple spin-down could not be done.


In order to address this problem, a program of regular monitoring of
the 5 confirmed AXPs by {\it RXTE} was initiated in 1998.  The goal was
to accomplish phase-coherent timing, in which every rotation of the
neutron star is counted on time scales of months to years.
Phase-coherent timing is done regularly for radio pulsars, and is
effective with any periodic source in which the periodicity is very
stable, or at least changes relatively slowly.  This turned out to
apply nicely to the AXPs \citep{kcs99}.  
For example, the RMS phase residual for 1E~2259+586 in $\sim$5~yr of
timing (pre-June 2002) is
under 2\% of the pulse period, following the removal of a model having
only three free parameters \citep[][hereafter GK02]{gk02}.
Phase-coherent timing on long time scales has now been accomplished for
AXPs RXS J1708$-$4009 \citep{kcs99}, 4U~0142+61 (GK02) and
1E~1841-045 \citep{ggk+02} and indicates these sources are capable of
great rotational stability.  This stability argues against an accretion
origin of the X-rays, since most accreting sources show much higher
levels of torque noise \citep[but see][]{bia+01}.  The stability is
comparable in some cases (particularly 4U~0142+61 and 1E~2259+586) to
that seen in young radio pulsars.  Together with the much noisier
timing properties of SGRs \citep[e.g.][]{wkv+99}, this provides support
for a continuum of timing noise properties in the radio pulsar, AXP and
SGR populations, in line with the magnetar model.

\begin{figure}[htb]
\includegraphics*[width=2.9in]{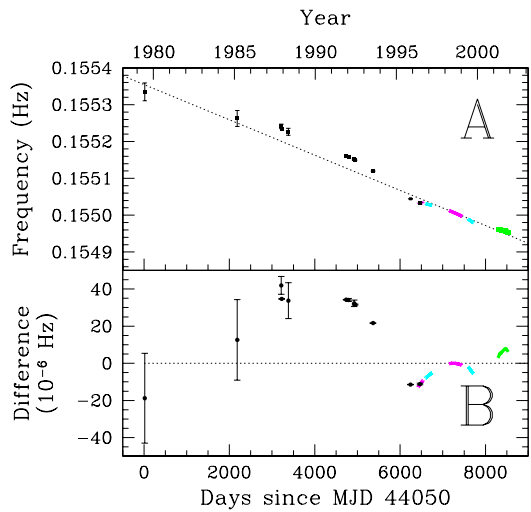}
\vspace{-0.4in}
\caption{Top: Long-term frequency history of 1E~1048$-$5937 (after
Kaspi et al. 2001).  The heavy lines represent intervals over which
phase-coherent timing was possible.  Bottom:  Same data but with
the linear trend removed.}
\label{fig:1048}
\end{figure}

However, one AXP, 1E~1048$-$5937, is a much noisier rotator than the
others, so much so that phase-coherent timing cannot be accomplished
over more than a few months at a time \citep{kgc+01}.  More
detailed observations of the source reveal that its spin-down rate can
change on time scales of weeks, and by large factors
(see Fig.~\ref{fig:1048}; Gavriil \& Kaspi, 
in preparation).  This behavior is reminiscent of that seen in SGRs
1806$-$20 and 1900+14 \citep{wkg+02}.

Thus, deviations from simple spin-down in AXPs appears to come in three
flavours:  (i) glitches and subsequent recovery; (ii) low-level
stochastic variations having a ``red'' spectrum, similar to the
``timing noise'' seen in radio pulsars; and (iii) large,
short-time-scale variations which preclude phase connection.   The
origin of the latter two in particular is unknown.  The low-level
variations in radio pulsars may be related to crustal superfluid
effects such as ``mini-glitches'', or may, in some cases, result from
long-term recoveries from glitches that preceded the commencement of
the observations.  \citet{act03} have recently suggested that 
the larger-scale torque variations arise from angular
momentum transfer from a superfluid core.  Such a core, they argue, also
results in a reduction in the interior temperature that could make the
crust more brittle, hence result in greater burst activity as seen in
the SGRs (and possibly 1E~1048$-$5937; see \S\ref{sec:bursts}).

\subsection{Glitches}

\begin{figure}[htb]
\includegraphics*[width=2.9in]{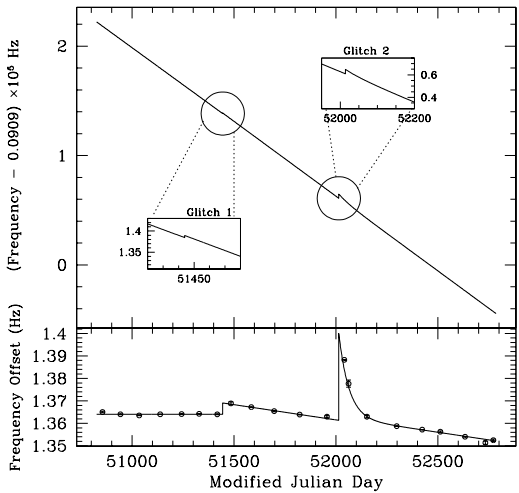}
\vspace{-0.4in}
\caption{Frequency history of RXS~J1708$-$4009 showing the two
very different glitches.  The top plot shows the overall frequency evolution,
while the bottom plot shows the same but with the long-term spin-down trend
removed, as well as measured frequencies.  The best-fit model is based
on a phase-coherent analysis (after Kaspi \& Gavriil 2003).}
\label{fig:1708glitches}
\end{figure}

Because phase-coherent timing counts every rotation, it determines spin
parameters with high precision.  This permits sensitivity to
glitches having fractional amplitudes as low as $\sim
10^{-7}$.  The first AXP glitch was detected in RXS J1708$-$4009
\citep{klc00}, and had fractional amplitude $6 \times 10^{-7}$, and an
increase in the magnitude of the spin-down rate of $\sim$1\%.  These
glitch properties are similar to those seen in Vela-like radio pulsars.
Interestingly, this source glitched again $\sim$1.5~yr later
\citep{kg03,dis+03}.  However, the second glitch was much larger, with
fractional frequency change $4 \times 10^{-6}$, and a significant
post-glitch recovery in which nearly all of the glitch relaxed on a
time scale of $\sim$50~days.  The frequency history of this source is
shown in Figure~\ref{fig:1708glitches}.  
Neither glitch was accompanied by any obvious radiative changes.

\begin{figure}[htb]
\includegraphics*[width=2.9in]{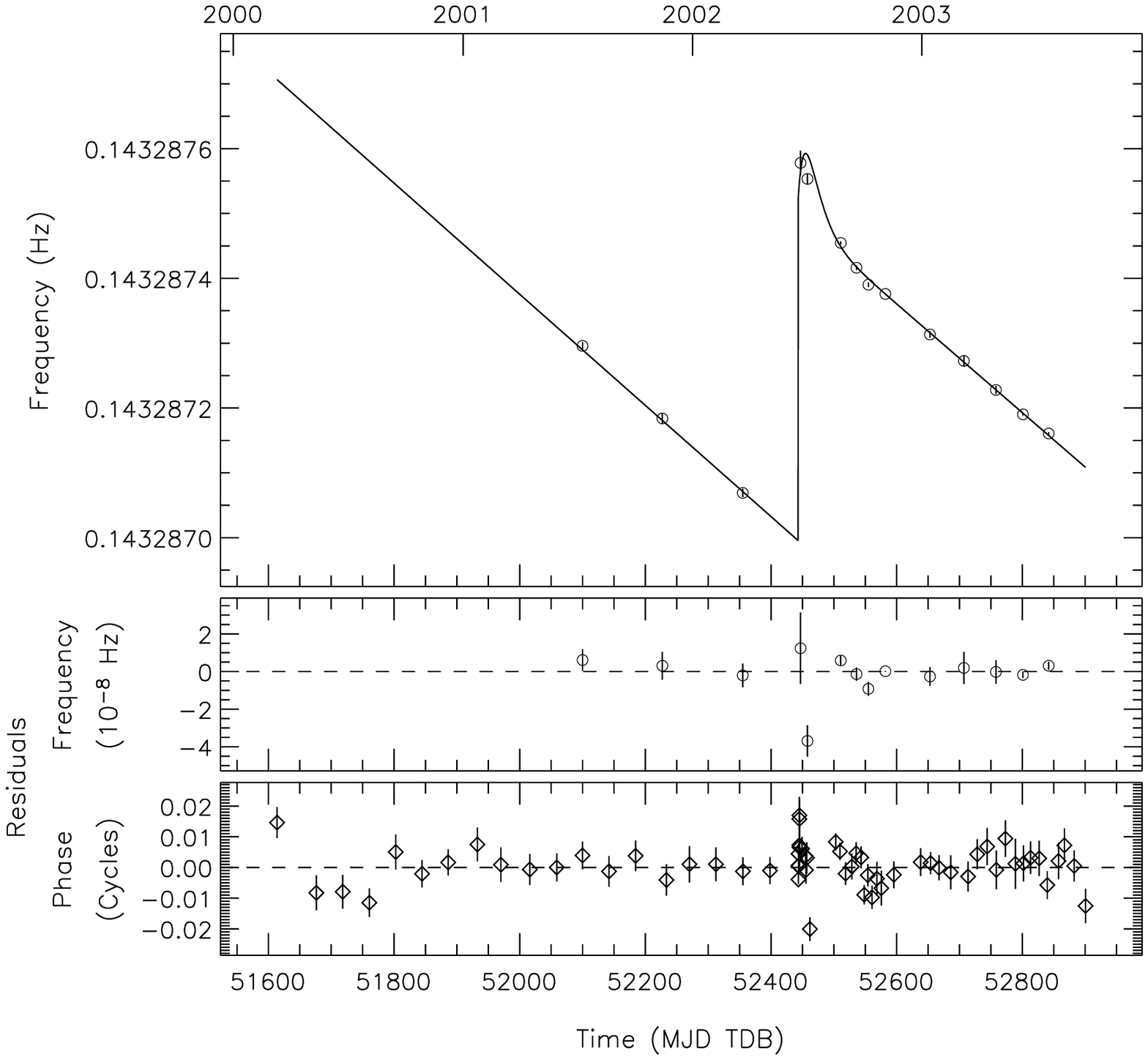}
\vspace{-0.4in}
\caption{Frequency history of 1E~2259+586 around the time of its 2002 outburst based
on a phase-coherent analysis.
The top plot shows the frequency evolution around the glitch,
along with measured frequences.
The middle panel shows frequency residuals, while the bottom shows phase residuals
(after Woods et al. 2003).}
\label{fig:2259glitch}
\end{figure}

The second discovered AXP glitch was in 1E~2259+586 \citep{kgw+03,wkt+03}.
Unlike the RXS~J1708$-$4009 glitches, this one
occurred simultaneously with (or possibly a few hours before -- see
Woods et al. 2003) a major outburst in which over 80 X-ray bursts were
detected in just a few hours, in addition to sudden order-of-magnitude
increases in the pulsed and unpulsed flux, significant pulse profile
changes, and an infrared enhancement (all discussed below).  This
represents the first neutron-star glitch ever observed to be accompanied by
significant radiative changes, and clearly indicates a major event that
simultaneously affected both the internal and external structure of the
star.  Roughly 20\% of the glitch recovered on a time scale of weeks,
and in doing so resulted in the stellar spin-down being a factor of
$>2$ greater than its pre-outburst value (Fig.~\ref{fig:2259glitch}).
This is unprecedented in radio pulsars, and suggests that just
following the glitch, the neutron star superfluid was actually spinning
{\it slower} than the crust, with the observed subsequent enhanced
spin-down a result of angular moment transfer from the crust back to
the superfluid (Woods et al. 2003).  Additionally, there was evidence
that the glitch may have been resolved in time, on a time scale of
$\sim$2 weeks.  

Glitches are definitely expected in the magnetar model \citep[e.g.][]{td96a}.
As pointed out by \citet{klc00}, at least in principle, an
accreting source can undergo a spin-up glitch since the latter
results from an internal angular momentum transfer from superfluid
to crust regardless of the nature of the external spin-down torque.
However, one would not expect simultaneous bursts in an accretion
scenario, as one might in the magnetar model.

The observed AXP glitches provide a very plausible explanation for the
historically observed spin-down deviations at least in 1E~2259+586
(Baykal \& Swank 1996).  The picture is still not clear for
1E~1048$-$5937, however.

\section{AXP X-RAY PULSE PROFILES AND PULSED FRACTIONS}
\label{sec:profiles}

AXP pulse profiles are, like those of the SGRs, broad, with large
(\gapp 80\%) duty cycles, and generally significant harmonic content
(e.g. GK02).  The profiles show energy dependences that vary from
source to source.  A possible trend of greater energy dependence for
profiles with higher harmonic content was identified by GK02, who also
showed that in general, AXP pulse profiles are very stable.

\begin{figure}[htb]
\includegraphics*[width=2.9in]{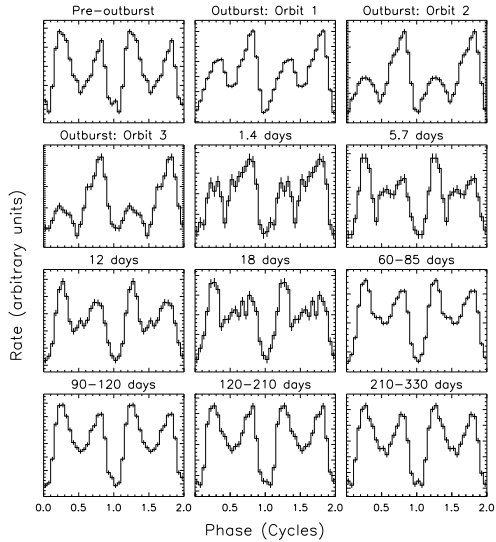}
\vspace{-0.4in}
\caption{Pulse profile changes in 1E~2259+586 seen around the time of the 2002 June 
outburst (after Woods et al. 2003).}
\label{fig:glitches}
\end{figure}

However, in 2002 June, simultaneously with the detection of the glitch
and X-ray bursts, the pulse profile of 1E~2259+586 underwent
significant changes, on time scales from hours to days (Kaspi et al.
2003; Woods et al. 2003).  The profile had relaxed back to its
pre-outburst morphology by $\sim$2 weeks following the outburst.  
\citet{ikh92} observed an apparent change
in the X-ray pulse profile of 1E~2259+586, in which the relative
amplitude of the two peaks in the profile changed between observations
made in 1989 and 1990.  This can be explained as being
due to an outburst having occurred just before the 1990
observation (see \S\ref{sec:fluxes}).

AXP pulsed fractions vary from source to source, with the highest being
$\sim$0.8 for 1E~1048$-$5937 and the lowest being $\sim$0.1 for
4U~0142+61.  Some, but not all, are energy dependent, and those which
are vary differently with energy.  For a summary of AXP pulsed
fractions and their energy dependences, see \citet{opk01}.
It is not clear whether the pulsed fractions are time variable
in general.  However, the pulsed fraction of 1E~2259+586 clearly
changed at the time of its 2002 outburst: immediately post-outburst,
the pulsed fraction decreased from its quiescent level of 
$\sim$0.23 to $\sim$0.15, however it recovered fully after 3 days
(Woods et al. 2003).

\section{X-RAY SPECTRA}

X-ray spectra of AXPs generally require two components.  These are
usually taken to be a thermal blackbody component with a power-law
tail.  The measured spectral parameters of the known AXPs are given in
Table~\ref{ta:axps2}.  The spectra as a class are softer than those of
the SGRs in quiescence.  The softest source in that class is
SGR~0525$-$66; its spectral parameters are actually softer than those
of 1E~1048$-$5937, which, among other things, prompted \citet{kkm+03}
and Kaspi et al.  (2001) to suggest these sources may be transition
objects between the two classes.

In the context of the magnetar model, the spectra can be understood as
follows.  The thermal component is emerging from the stellar surface, a
result of heating of the interior by active magnetic field decay
\citep{td95,td96a}.
The thermal spectrum is thought to deviate significantly from that of a
blackbody, because of the effects of the stellar atmosphere,
as well as the large magnetic field, which results
in different opacities for different photon polarizations, as well
as on QED vacuum polarization \citep{hl01a,oze01,ztst01,hl03,oze03}.
The thermal spectrum is hardened relative to a
blackbody of the same temperature due to the non-grey
atmosphere, although vacuum polarization counteracts this slightly.  As
observers fit the thermal component with a blackbody, some portion of
the non-thermal component may result from the atmospheric distortion.
However, this portion is probably small.  A more promising
origin of the non-thermal emission is external resonant Compton
scattering of thermal seed photons by magnetospheric currents \citep{tlk02}.

The X-ray spectra were, pre-{\it Chandra} and {\it XMM-Newton}, hoped
to hold direct evidence for the high magnetic field via features such
as electron cyclotron lines \citep[e.g.][]{hl01a,ztst01,oze03,hl03}.
Of course an electron cyclotron line in a $\sim 10^{15}$~G field might
look similar to a proton cyclotron line in a $\sim 10^{12}$~G field.
In any case, no such lines have been seen in spite of some high
spectral, and in some cases, temporal resolution observations
\citep{pkw+01,jmcs02,tgsm02,wkt+03}.

\section{AXP X-RAY FLUX VARIABILITY}
\label{sec:fluxes}

Strong flux variability pre-{\it BeppoSax} and {\it RXTE} was
reported for 1E~2259+586 and 1E~1048$-$5937 (Baykal \& Swank 1996;
Oosterbroek et al. 1998).  Flux variations of a factor of 5--10 were
reported, albeit from different instruments, having different spectral
responses, with some imaging and some not.  However Iwasawa et al.
(1992) reported a brightening of a factor of $\sim$2 in a 1990 {\it GINGA}
observation of 1E~2259+586 compared with an observation in
1989.  They noted that the 1990 pulse profile was also significantly
different than that observed previously, with different relative peak
amplitudes, and different peak shapes.  Furthermore, the measured 1990
spin period was fractionally shorter by $\sim 3 \times 10^{-6}$
compared with what the previous spin-down rate would have predicted.

In $\sim$5~yr of monitoring using the PCA on {\it RXTE}, GK02
found no evidence for such flux variations in any AXP.  This was
consistent with what was found by Tiengo et al. (2002) in a comparison
of past observations of 1E~1048$-$5937 with recent {\it XMM-Newton}
data.  The overall recent lack of variability in AXPs thus appeared
discrepant with the historical record.

\begin{figure}[htb]
\includegraphics*[width=2.9in]{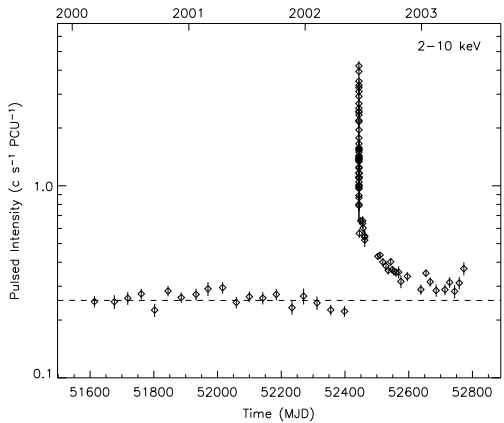}
\vspace{-0.4in}
\caption{The pulsed flux history of 1E~2259$+$586 (2$-$10 keV) around
the time of the outburst, indicated by the sharp spike (after Woods et al. 2003).}
\label{fig:2259flux}
\end{figure}

The 2002 June outburst of 1E~2259+586 appears to have solved this
conundrum, at least for this source.  Simultaneous with the bursting
were increases in the and persistent fluxes by a factor of $>10$ (Kaspi
et al. 2003; Woods et al. 2003), which mostly decayed on a time scale
of days, but which has left an X-ray afterglow in which the pulsed flux
is still a factor of $\sim$2 greater than the pre-outburst value a year
since the outburst (Fig.~\ref{fig:2259flux}).  The total energy in
excess pulsed and persistent emission during the short-decay-time-scale
enhancement was 3 $\times$ 10$^{39}$~erg (2--10~keV), while that in the
extended afterglow is much more, $2\times 10^{41}$~erg (Woods et al.
2003).

The rapidly decaying flux enhancement seen in 1E~2259+586 could be due
to a transient surface hot spot.  During the rapid initial flux decay,
the blackbody radius was smaller than at all other times, the
temperature was higher, and the pulse profile was clearly different,
supporting this picture.  Alternatively, it could have been
magnetospheric, as a large current density will be excited in the
magnetosphere above regions of strong crustal shear.  The short-lived
afterglows detected after intermediate SGR bursts have a simple
explanation as the cooling of a pair-rich surface layer heated by a
high-energy flare \citep{isw+01}.  However, no such flare was seen for
1E~2259+586.  This is problematic also for explaining the
long-time-scale afterglow.  In SGRs, bulk heating of the crust can
power an excess heat flux from its surface for a year or more, and has
been proposed as the explanation for the quasi-power-law flux decay
seen in SGR~1900$+$14 \citep{let02} and SGR~1627$-$41 \citep{kew+03}.
In each case, an initial deposition of 10$^{44}$~ergs was assumed,
consistent with the detection of an initial giant soft gamma-ray flare;
this was unseen for 1E~2259+586.  For a more detailed discussion of the
possible origins of the enhanced emission, see Woods et al. (2003).

We note that the combination of the observed flux enhancement, glitch
and pulse profile change in 1E~2259+586 observed by Iwasawa et al.
(1992) using {\it GINGA} are consistent with an outburst similar to
that observed in 2002 June having occurred days/weeks prior to their
1990 observation (see \S\ref{sec:profiles}).  This offers an estimate
of a crude burst rate of two every $\sim$20~yr.

This observed variability associated with outbursts also makes the two
transient AXP candidates (see Table~\ref{ta:axps1}), AX~J1845$-$0258
\citep{vg97}, and XTE~J1810$-$197 \citep{ims+03} easier to understand.
These two objects have both shown factor of $>10$ increases in their
fluxes.  For XTE~J1810$-$197, the flux decreased slowly after its
appearance, in concert with its spin-down rate, not unlike the behavior
seen in 1E~2259+586 post-outburst (Ibrahim et al. 2003; Woods et al.
2003).   Such transient AXPs suggest a large population of quiescent
AXPs exists in the Galaxy.

\section{X-RAY BURSTS}
\label{sec:bursts}

The first discovery of bursts from AXPs came from the {\it RXTE}/PCA
monitoring observations of 1E~1048$-$5937.  Two faint bursts, separated by
$\sim$2~weeks, were detected in $\sim$425~ks of exposure over $\sim$5~yr
\citep{gkw02}.  These bursts very much resemble SGR
bursts.  Specifically, their fast rise times, short durations, hard
spectra relative to the quiescent emission, fluence and probably
clustering, are all SGR burst hallmarks.  The origin of the
bursts could not unambiguously be proven to be the AXP, given the large
PCA field-of-view, and the absence of any other radiative or spin
change in the source.  Intriguingly, the first burst's spectrum was not
well fit by a continuum model, showing evidence for a strong emission
line at $\sim$14~keV.

Not long after the reporting of the above two bursts, a major outburst
consisting of over 80 bursts was detected from the direction of
1E~2259+586 fortuitously during a regular {\it RXTE}/PCA monitoring
observation in 2002 June (Kaspi et al. 2003).  These bursts were
very similar to those of SGRs \citep{gkw03}.
Specifically, like the SGRs, the AXP burst
durations follow a log-normal distribution which peaks at 99~ms, the
differential burst fluence distribution is well described by a power
law of index $-1.7$, the burst fluences
are positively correlated with the burst durations, the distribution of
waiting times is well described by a log-normal distribution of mean
47~s, and the bursts are generally asymmetric with faster rise than
fall times.


However, there were some notable differences between the AXP and SGR
bursts that may be clues to the physical differences between the two
source classes (Gavriil et al. 2003).  Specifically, the AXP
bursts exhibit a wider range of durations and, unlike SGR bursts, occur
preferentially near pulse maxima; the correlation between burst fluence
and duration seen for SGRs is flatter than for SGRs; the AXP bursts are
on average less energetic than are SGR bursts; and the more energetic
AXP bursts have the hardest spectra -- the opposite of what is seen for
SGRs (Gavriil et al. 2003).  Furthermore, in stark
contrast to SGRs, the energy detected in bursts ($6 \times 10^{37}$
erg, 2--60~keV) was much smaller than that in the post-outburst
persistent flux enhancement ($2\times 10^{41}$~erg, 2--10~keV).
This could indicate bursting activity that was missed by our
observations and the gamma-ray monitors, although the latter would have
easily detected SGR-like bursts having the missing energy (Woods et al.
2003).  This ``quiet'' outburst strongly suggests there are many more
such objects in the Galaxy than was previously thought, as is also
indicated by the transient AXP candidates (see \S\ref{sec:fluxes}).

Overall, the properties of the outburst in 1E~2259+586 argue that the
star suffered a major event that was extended in time and had two
components, one tightly localized on the surface of the star (i.e. a
fracture or a series of fractures) and the second more broadly
distributed (possibly involving a smoother plastic change).  The glitch
points toward a disturbance within the superfluid interior while the
extended flux enhancement and pulse profile change suggest an
excitation of magnetospheric currents and crustal heating. The very
rich data set provided by this outburst should be very useful in
constraining physical properties of the affected neutron star.

\section{OPTICAL/IR OBSERVATIONS AND VARIABILITY}

Of the five confirmed AXPs, four now have secure or possible optical/IR
counterparts.  The first optical/IR detection of an AXP was made of
4U~0142+61, by \citet{hvk00}.  They argued that the source, which had
$R\simeq 25$~mag, was too dim to be from an accretion disk.  This was
confirmed by \citet{km02} who showed that this source is pulsing with
the X-ray period and a 27\% pulsed fraction, much too high to be
reprocessed light.  \citet{htvk01} discovered a possible near-IR ($K_s
= 21.7$~mag) counterpart to 1E~2259+586.  This was confirmed by Kaspi
et al. (2003) who found the source to have brighted by a factor of
$\sim$3 three days after its 2002 outburst, but faded by a factor of
$\sim$2 1~week later.  This source appears to have stayed brighter than
pre-outburst, however, two months after the outburst \citep{isc+03}.
\citet{wc02} reported a likely near-IR counterpart to 1E~1048$-$5937
having $K=19.4$~mag, and detected in multiple wavebands.
\citet{ics+02} reported significant IR variability in the source on a
time scale of $\sim$50~days.  They speculated the variability might be
related to bursting activity from this source (Gavriil et al. 2002), in
analogy with the variability seen in 1E~2259+586.  \citet{icp+03}
reported the detection of a possible IR counterpart to
RXS~J1708$-$4009.

These detections are interesting for two reasons.  First, as noted
above, in two cases variability is observed; for 1E~2259+586, it is
likely that it is associated with its 2002 June outburst.  Such
variability may therefore prove to be an important observational model
constraint, although currently its origin in the magnetar model is
unclear.  Most likely the optical/IR emission is a product (or, given
the large X-ray to optical/IR luminosity ratios, $L_x/L_{\rm IR} >
500$, a byproduct) of radiation processes in the outer magnetosphere,
and therefore is sensitive to the changes in the current structure
induced by magnetic reconfigurations.  Continued monitoring for
correlated optical/IR and torque variations seems warranted, especially
as a way of testing the proposed ``twisted magnetosphere'' model of
magnetars (Thompson, Lyutikov \& Kulkarni 2002).  Second, as pointed
out by \citet{icp+03}, the spectral energy distributions of the
optical/IR and X-ray emission show that the former is much too bright
to be the simple extrapolation of the blackbody component of the X-ray
spectrum.  However, it is fainter than the extrapolation of
the X-ray power-law spectral component, so calling it an ``excess''
may be premature.

\section{CONCLUSIONS AND OPEN QUESTIONS}

Since the 1996 commissioning of {\it BeppoSAX} and {\it RXTE}, our
overall picture of AXPs has changed dramatically.  The number of likely
AXPs has nearly tripled, and our understanding of these unusual
sources' properties has improved tremendously.  Perhaps the single most
important discovery is that the apparent resemblance of AXPs with SGRs
noted by Thompson \& Duncan in 1995 is more than skin deep: with the
discovery of bursts from AXPs, the two source classes are now united 
unambiguously.  Our next challenge is to learn how to extract
physically interesting information from AXP and SGR observations.  In
this sense, their study is still in its infancy, with observations
ahead of theory.

This said, there are obvious possibilities for fruitful observational
investigation of AXPs. First, it would be nice to have a direct
measurement of the inferred high magnetic field.  As no X-ray spectral
features have been forthcoming, another avenue is needed.  X-ray
polarization observations are an excellent possibility, particularly
for the brightest AXPs.  
The detection of such polarization, in addition to
confirming the high magnetic field, would be the first demonstration of
the birefringence of the vacuum, as predicted by QED.  In the shorter
term, glitches in AXPs may offer a practical method of constraining the
structure and physics of these objects.  The simulteneity of the
1E~2259+586 glitch with its outburst and associated radiative changes,
we suspect, is telling us a lot about the stellar structure.  Continued
patient timing of these objects has the potential to reveal
correlations between glitch properties like amplitude and relaxation
time scales with radiative properties, which will help us understand
properties of the highly magnetized crust and superfluid interior.
Optical/IR observations also offer hope of constraining magnetar outer
magnetosphere processes, although its origin is not yet clear.
Finally, there is the open question of the radio pulsar/AXP
connection.  Recently, several radio pulsars having inferred magnetic
fields {\it higher} than that of 1E~2259+586 have been discovered, yet
with no evidence for any AXP-like X-ray emission
\citep[e.g.][]{pkc00,msk+03}.  This is puzzling.  It may simply reflect
the fact that the magnetic field measured by $P$ and $\dot{P}$ is
approximate only, in which case the discovery of more such radio
pulsars should eventually result in the identification of the ``missing
link.''

\bigskip
Funding for this work comes from NSERC (via a Discovery Grant and
Steacie Supplement), the Canada Research Chair Program, NATEQ (via
Team and Observatoire de Mont Megantique Grants), CIAR (via a
Fellowship), and NASA (via the LTSA program).


\end{document}